\documentclass[twocolumn,aps,showpacs]{revtex4}
\usepackage{amssymb}
\usepackage{amsmath,bm}
\usepackage{graphicx}
\usepackage[normalem]{ulem}

\usepackage[usenames, dvipsnames]{xcolor}
\usepackage{lipsum}

\renewcommand{\sout}{\bgroup \color{red} \ULdepth=-.5ex \ULset}

\begin{document}
\title{Effects of energy conservation on equilibrium properties  of hot asymmetric nuclear matter}

\author{Zhen Zhang\footnote{zhenzhang$@$comp.tamu.edu}}
\affiliation{Cyclotron Institute and Department of Physics and Astronomy, Texas A$\&$M University, College Station, Texas 77843, USA}
\author{Che Ming Ko\footnote{ko$@$comp.tamu.edu}}
\affiliation{Cyclotron Institute and Department of Physics and Astronomy, Texas A$\&$M University, College Station, Texas 77843, USA}
\date{\today}

\begin{abstract}
Based on the relativistic Vlasov-Uehling-Uhlenbeck transport model, which includes relativistic scalar and vector potentials on baryons, we consider a $N-\Delta-\pi$ system in a box with periodic boundary conditions  to study the effects of energy conservation in particle production and absorption processes on the equilibrium properties of the system.  The density and temperature of the matter in the box are taken to be similar to the hot dense matter formed in heavy ion collisions at intermediate energies. We find that to maintain the equilibrium numbers of $N$, $\Delta$ and $\pi$, which depend on the mean-field potentials of $N$ and $\Delta$, requires the inclusion of these potentials in the energy conservation condition that determines the momenta of outgoing particles after a scattering or decay process.  We further find that the baryon scalar potentials mainly affect the $\Delta$ and pion equilibrium numbers, while the baryon vector potentials have considerable effects on the effective charged pion ratio at equilibrium. Our results thus indicate that it is essential to include in the transport model the effect of potentials in the energy conservation of a scattering or decay process, which is ignored in most transport models, for studying pion production in heavy ion collisions. 
\end{abstract}

\maketitle

\section{Introduction}

The charged pion ratio in intermediate energy heavy ion collisions has attracted much attention since it was proposed as a potential probe of the high-density behavior of nuclear symmetry energy~\cite{Li2002}. The latter is essential for understanding the properties of neutron  stars and gravitational waves from spiraling neutron star binary, but is still poorly known~\cite{Steiner2005b,Lattimer2007a,Li2008,Fattoyev2014}.  Various transport models have been used to constrain the nuclear symmetry energy at supra-saturation densities from experimental data on the $\pi^-/\pi^+$ ratio, but these studies have led to conflicting conclusions~\cite{Xiao2009,Feng2010,Xie2013}.  Therefore, more in-depth studies and careful modeling of pion production in heavy ion collision are important, especially because more systematic experimental measurements of the pion yield from intermediate energy heavy ion collisions are being carried out by the FRIB-RIKEN Spirit Collaboration in Japan~\cite{SEP}.  A lot of efforts have already been made recently to study various effects on pion production in heavy ion collisions at near-threshold energies, e.g., the pion in-medium potentials~\cite{Xu2010, Xu2013,Cozma2017,Zhang2017}, isovector potential of $\Delta$~\cite{Li2015}, threshold effects~\cite{Ferini2005,Song2014}, neutron-skin thickness~\cite{Wei2014}, and nucleon short-range correlation~\cite{Li2015a}.

As to the mean-field potential of $\Delta$ resonances in nuclear medium~\citep{Li2015, Cai2015}, it is commonly assumed in transport models that its isoscalar part is the same as that for nucleon, but its isovector part is taken to be the weighted average of those for neutron and proton according to the squared Clebsch-Gordan coefficients from its isospin structure~\cite{Li2002,Li2002b}. This assumption leads to a change between the initial and final potentials in some of the processes $N+N\leftrightarrow N+\Delta$ and $\Delta \leftrightarrow N+\pi$.  In essentially all transport studies except Refs.~\cite{Song2014,Cozma2016}, this potential difference has been neglected in the energy conservation condition for the above processes.  As shown in Ref.~\cite{Cozma2016} based on a non-relativistic quantum molecular dynamic model, this would lead to a violation of the local energy of scattering particles as well as the total energy of the system if the mean-field potentials are momentum dependent.  Besides affecting  the threshold energies for $\Delta$ and pion production as discussed in Refs.~\cite{Ferini2005,Song2014}, neglecting the change of potentials in these processes can also lead to an incorrect charged pion ratio.  To demonstrate this effect more transparently, we employ in this work the relativistic Vlasov-Uehling-Uhlenbeck (RVUU) transport model~\cite{Ko1987,Ko1988,Ko1996,Song2014} to show that to obtain the correct equilibrium numbers of $N$, $\Delta$ and $\pi$ in a box with periodic boundary conditions requires the inclusion of the potentials in the energy conservation condition for determining the momenta of final particles in the processes $N+N\leftrightarrow N+\Delta$ and $\Delta \leftrightarrow N+\pi$. The temperature and density of the hadronic matter in the box are taken to be similar to those formed in the high density stage of heavy ion collisions at intermediate energies~\cite{Xu2013}.  We find that to retain the equilibrium $N$, $\Delta$ and $\pi$ numbers, it is necessary to take into account the mean-field potentials in the energy conservation conditions for above processes. In particular, we find that the vector potential has considerable effects on the effective charged pion ratio, while the scalar potential mainly affects the total pion and $\Delta$ numbers.  Our results thus demonstrate the importance of treating in heavy ion collisions the effects of mean-field potentials not only on particle propagation but also on their scatterings. The present study complements the recent transport code comparison project~\cite{Xu:2016lue} to improve the robustness of transport model predictions for collisions of neutron-rich nuclei at low and intermediate energies. 

The paper is organized as follows. In Sec. II, we introduce the baryon mean-field potentials in isospin-asymmetric matter and the detailed balance relations for particle scatterings in medium. Results on $\Delta$ and pion numbers and their momentum spectra are shown and discussed in Sec. III.  Finally, we give a summary in Sec. VI. 

\section{The model }

\subsection{Baryon mean field potentials in isospin-asymmetic  matter}
\label{Sec:RMF}

In the present work, we focus on the effects of baryon potentials and treat pions as if they are in free space. The mean-field potentials of nucleons and $\Delta$ resonances are taken from the nonlinear relativistic mean field NL$\rho$ model~\citep{Liu2002}. In this model, the Lagrangian density is given by 
\begin{eqnarray}
\mathcal{L}&=&\bar{N}[\gamma_{\mu}(i\partial^{\mu} - g_{\omega}\omega^{\mu}
-g_{\rho}\bm{\tau}\cdot\bm{\rho}^{\mu})-(m_N-g_{\sigma}\sigma)] N  \notag\\
& & +\frac{1}{2}\left( \partial_{\mu}\sigma\partial^{\mu}\sigma
-m_{\sigma}^2\sigma^2\right)-\frac{a}{3}\sigma^3-\frac{b}{4}\sigma^4  \notag\\
& &-\frac{1}{4}\Omega_{\mu\nu}\Omega^{\mu\nu}
+\frac{1}{2}m_{\omega}^2\omega_{\mu}\omega^{\mu} \notag \\
& &-\frac{1}{4}\bm{R}_{\mu\nu}\cdot \bm{R}^{\mu\nu} +\frac{1}{2}m_\rho^2\bm{\rho}_{\mu}\cdot\bm{\rho}^{\mu}.
\end{eqnarray}
In the above, $N$ denotes the nucleon field, and $\sigma$, $\omega_{\mu}$ and $\bm{\rho}_{\mu}$ represent the isoscalar-scalar, isoscalar-vector and isovector-vector meson fields, respectively.  The antisymmetric field tensors $\Omega$ and $\bm{R}_{\mu\nu}$ are given by   $\Omega_{\mu\nu}=\partial_{\mu}\omega_{\nu}-\partial_{\nu}\omega_{\mu}$ and $\bm{R}_{\mu\nu}=\partial_{\mu}\bm{\rho}_{\nu}-\partial_{\nu}\bm{\rho}_{\mu}$, respectively. For the $\Delta$ resonance, it is considered as a molecular state of nucleon and pion, and its interactions with mesons can be treated as the weighted average of those for a neutron and a proton based on its isospin structure~\citep{Song2014}.  For example, the coupling of a $\Delta^+$ to a meson is given by $2/3$ of that for the proton plus $1/3$ of that for the neutron.

In the mean-field approximation, with the inclusion of $\Delta$ resonances, the meson field equations have the following form:
\begin{eqnarray}
\partial_{\nu}\partial^{\nu}\sigma + m_{\sigma}^2\sigma+a\sigma^2+b\sigma^3 &=& g_{\sigma}\phi_B, \label{Eq:sigma}\\ 
(\partial_{\nu}\partial^{\nu}+m_{\omega}^2)\omega^{\mu}&=&g_{\omega}j_B^{\mu},  \\ \label{Eq:omega}
(\partial_{\nu}\partial^{\nu}+m_{\rho}^2)\rho_3^{\mu}&=&g_{\rho}j_I^{\mu}\label{Eq:rho},
\end{eqnarray}
where $\phi_B$, $j_B^{\mu}$ and $j_I^{\mu}$ are the baryon scalar, vector and isovector vector densities given by 
\begin{eqnarray}
\phi_B&=&\phi_p+\phi_n +\phi_{\Delta^{++}}+\phi_{\Delta^+}+\phi_{\Delta^0}+\phi_{\Delta^-},  \\
j_B^{\mu}&=&j^{\mu}_p+j^{\mu}_n +j^{\mu}_{\Delta^{++}}+j^{\mu}_{\Delta^+}+j^{\mu}_{\Delta^0}+j^{\mu}_{\Delta^-} , \\
j_I^{\mu}&=&j^{\mu}_p-j^{\mu}_n +j^{\mu}_{\Delta^{++}}+j^{\mu}_{\Delta^+}/3-j^{\mu}_{\Delta^0}/3-j^{\mu}_{\Delta^-}.
\end{eqnarray} 
In the above, the scalar and vector densities are defined as
\begin{eqnarray}
\phi_i&=&\int\frac{d^3\bm{p}_i}{(2\pi)^3}\frac{m_i^{\ast}}{p_0^{\ast}}f_i(\bm{p}_i) ,\label{Eq:rhos}\\
j_i^{\mu}&=&\int\frac{d^3\bm{p}_i}{(2\pi)^3}\frac{p_i^{\mu\ast}}{p_0^{\ast}}f_i(\bm{p}_i),
\end{eqnarray}
where $i= p,n, \Delta^{++},\Delta^+,\Delta^0,\Delta^-$, the $f_i(\bm{p})$ is the baryon distribution function including the spin degeneracy, and the effective mass $m_i^*$ and kinetic energy-momentum $p_i^{\mu*}$ are defined by
\begin{eqnarray}
m_i^{\ast} & = &  m_i-g_{\sigma}\sigma, \\
p_i^{\mu*} & = &  p_i^{\mu}-g_{\omega}\omega^{\mu}-x_ig_{\rho}\rho^{\mu}\label{Eq:pstar}.
\end{eqnarray}
Here $p_i^{0*}= \sqrt{m_i^{*2}+\bm{p}_i^{*2}}$, and  $x_p=1, x_n=-1 , x_{\Delta^{++}}=1, x_{\Delta^+}=1/3  , x_{\Delta^{0}}=-1/3  , x_{\Delta^{-}}=-1$. We note that the time 
component of the vector density $j_i^{\mu}$ is just the number density of baryon $i$. 

For static and uniform hadronic matter in a box, all derivative terms and spatial components of vector densities in Eqs.~(\ref{Eq:sigma})-(\ref{Eq:rho}) are zero. The vector potential is then completely determined by the baryon density $\rho =\rho_p+\rho_n +\rho_{\Delta^{++}}+\rho_{\Delta^+}+\rho_{\Delta^0}+\rho_{\Delta^-}$ and the isovector density $\rho_I=\rho_p-\rho_n +\rho_{\Delta^{++}}+\rho_{\Delta^+}/3-\rho_{\Delta^0}/3-\rho_{\Delta^-}$ through the relations $m_{\omega}^2 \omega^0=  g_{\omega}\rho$ and $m_{\rho}^2 \rho_3^0=  g_{\omega}\rho_I$.  For baryon density $\rho = 0.24 \mathrm{fm}^{-3} (1.5\rho_0)$  and isospin asymmetry $\delta_{\mathrm{like}} =\rho_I/\rho = 0.2$, the resulting isoscalar-vector potential is  $g_{\omega}\omega^0 = 256.6$ MeV and the isovector-vector potential is $g_{\rho}\rho_3^0 = -7.0~\mathrm{MeV}$.
 
For simplicity, we neglect the quantum nature of $N$, $\Delta$ and $\pi$, and use the Boltzmann distribution to describe their equilibrium distributions. For pions and nucleons, their momentum distribution functions at a given temperature $T$ are then given by 
\begin{equation}
\label{Eq:fnpi}
f_i(\bm{p}_i) = g_i\mathrm{exp}\left[ -\frac{E_i-\mu_i}{T} \right],
\end{equation}
with $i= n,p, \pi^{+}, \pi^0, \pi^-$, the spin degeneracy $g_i = 1 (2)$ for pion (nucleon), and
$\mu_i$ and $E_i=p^0_i$ being the chemical potential and energy of particle $i$, respectively. For the $\Delta$ resonance, which has a mass distribution, its momentum distribution function is 
\begin{equation}
\label{Eq:fdlt}
f_i(\bm{p}_i)  = 4\int\frac{dm}{2\pi} \mathcal{A}(m)\mathrm{exp}\left[ -\frac{E_i-\mu_i}{T} \right],
\end{equation} 
where $i=\Delta^{++}, \Delta^+, \Delta^0, \Delta^-$, the factor 4 is the spin degeneracy, and $\mathcal{A}(m)$ is the normalized spectral function 
\begin{equation}
\mathcal{A}(m)=\frac{1}{\mathcal{N}}\frac{4\protect m_0^2\Gamma}{(m^2-m_0^2)^2+m_0^2\Gamma^2},
\end{equation}
with $m_0 =1.232~\mathrm{GeV}$ being the pole mass of a $\Delta$ resonance and $\mathcal{N}$ being the normalization factor. Taking the decay width of $\Delta$ in free space  to be ~\cite{Kitazoe1986}
\begin{equation}
\Gamma = \frac{0.47q^3}{(m_{\pi}^2+0.6q^2)},
\end{equation}
where $q$ is the pion momentum in $\Delta$ rest frame, gives the normalization factor  $\mathcal{N}=0.948$.  For a N-$\Delta$-$\pi$ system at thermal and chemical equilibrium, the chemical potentials of nucleons, pions and $\Delta$ resonances satisfy following relations:
\begin{eqnarray}\label{chemical}
\mu_{\Delta^{++}} & =& 2\mu_p-\mu_n, \nonumber\\ 
\mu_{\Delta^{+}} & =&  \mu_p,\nonumber\\ 
\mu_{\Delta^{0}} & =& \mu_n, \nonumber\\ 
\mu_{\Delta^{-}} & =& 2\mu_n-\mu_p, \nonumber\\ 
\pi_{\pi^{+}} & =& \mu_p-\mu_n, \nonumber\\ 
\mu_{\pi^{0}} & =& 0, \nonumber\\ 
\mu_{\pi^{-}} & =& \mu_n-\mu_p, 
\end{eqnarray}
Given the temperature $T$, baryon density $\rho_B$ and isospin asymmetry $\delta_{\mathrm{like}}$, the baryon mean-field potentials and numbers can be obtained by solving above equations iteratively.

\subsection{Detailed balance relations in nuclear medium}
\label{Sec:DB}

In the RVUU model, the cross sections for the inverse reactions $N+\Delta \rightarrow N+N$ and $N+\pi\rightarrow \Delta$ are related to those for the reactions $N+N \rightarrow N+\Delta$ and $\Delta \rightarrow N+\pi$ by detailed balance conditions, which guarantee the $N-\Delta-\pi$ system in the box to reach the correct equilibrium distributions.

For the reaction $1+2\rightarrow 3+4$ (where 1, 2, 3 are nucleons and 4 is $\Delta$), the total cross section is given by 
 \begin{eqnarray}
\label{Eq:sig12}
\sigma_{12} &=&\int \frac{dm}{2\pi}\mathcal{A}(m) \int \frac{d\bm{p}_3^*}{(2\pi)^3}\frac{d\bm{p}_4^{\ast}}{(2\pi)^3}\frac{1}{4E_3^*E_4^*} \notag\\
&\times&\frac{\vert\mathcal{M} \vert^2_{12\rightarrow 34}}{4E_1^*E_2^*\vert \bm{v}_1-\bm{v}_2\vert}  (2\pi)^4\delta^4(p_1+p_2-p_3-p_4),\notag\\
\end{eqnarray}
where $m$ is the mass of $\Delta$,  $\bm{p}_i^*$ and $E_i^*$ are the kinetic momentum and energy of particle $i$ (i=1,2,3,4), $p_i$ is its canonical four momentum, and $\vert \bm{v}_1-\bm{v}_2\vert = \vert \bm{p}_1^*/E_1^*-\bm{p}_2^*/E_2^*\vert$ is the relative velocity of particles 1 and 2.  Note that for the process $N+N\rightarrow N+\Delta$, since the total initial potential may be different from the total final potential, the initial and final kinetic energies and momenta are not necessarily conserved.  Evaluating the integral in Eq.(\ref{Eq:sig12}) in the frame $F^{\prime}$ of $\bm{p}_3^*+\bm{p}_4^*=0$, whose 
velocity is given by 
\begin{equation}
\beta = \frac{\bm{p}_3^*+\bm{p}_4^*}{E_3^*+E_4^*}=\frac{\bm{p}_1+\bm{p}_2-\bm{\Sigma}_3-\bm{\Sigma}_4}{E_1+E_2-\Sigma^0_3-\Sigma^0_4},
\end{equation}
with $\Sigma^0_i$ and $\bm{\Sigma}_i$ being the time and spatial components of the vector mean field of particle $i$, we obtain 
\begin{eqnarray}\label{Eq:sig12fb}
&&\sigma_{12}= \frac{1}{32\pi^2}\int dm \mathcal{A}(m)\int \frac{p_4^{*'2}dp_4^{*'}}{E_3^{*'}E_4^{*'}}\frac{\vert\mathcal{M}\vert^2_{12\rightarrow 34}}{4E_1^{*'}E_2^{*'}\vert \bm{v}_1^{'}-\bm{v}_2^{\prime}\vert} \notag \\
&&\times \delta(E_1^{*'}+E_2^{*'}+\Sigma^{0'}_1+\Sigma^{0'}_2-E_3^{*'}+E_4^{*'}+\Sigma^{0'}_3+\Sigma^{0'}_4), \notag \\
&&=\frac{1}{16\pi}\frac{p_4^{\ast '}}{E_3^{\ast '}+E_4^{\ast '}}\frac{\vert \mathcal{M} \vert ^2_{12\rightarrow 34}}{E_1^{\ast '}E_2^{\ast '}\left\vert \frac{\bm{p}_1^{\ast '}}{E_1^{\ast '}}-
\frac{\bm{p}_1^{\ast '}}{E_1^{\ast '}} \right\vert},
 \end{eqnarray}
where the prime indicates quantities in the frame $F'$.

Analogously, the cross section for the raction $3+4\rightarrow 1+2$ in the nuclear matter frame 
is
\begin{eqnarray}
\label{Eq:sig34fb}
\sigma_{34} = \frac{1}{16\pi}\frac{p_1^{\ast ''}}{E_1^{\ast ''}+E_2^{\ast ''}}\frac{\vert \mathcal{M} \vert ^2_{34\rightarrow 12}}{E_3^{\ast ''}E_4^{\ast ''}\left\vert \frac{\bm{p}_3^{\ast ''}}{E_3^{\ast '' }}-\frac{\bm{p}_4^{\ast '' }}{E_4^{\ast '' }} \right\vert},
\end{eqnarray}
where the double prime indicates quantities in the frame $F^{''}$ of $\bm{p}_1^{\ast}+\bm{p}_2^{\ast}=0$.  

Comparing Eqs.~(\ref{Eq:sig12fb}) and (\ref{Eq:sig34fb}) and using the relation $\vert\mathcal{M}\vert^2_{12\rightarrow34} =2\vert\mathcal{M}\vert^2_{34\rightarrow12} $, we obtain the following detailed balance relation: 
\begin{eqnarray}
\sigma_{34} &=& \frac{\sigma_{12}}{2(1+\delta_{12})}\frac{2\pi p_1^{\ast \prime\prime}}{\int dm \mathcal{A}(m)p_4^{\ast\prime}(m)} \nonumber\\ 
& &\times \frac{E_3^{\ast\prime}+E_4^{\ast\prime}}{E_1^{\ast\prime\prime}+E_2^{\ast\prime\prime}}\frac{\vert E_1^{\ast '}\bm{p}_2^{\ast '}-E_2^{\ast '}\bm{p}_1^{\ast '}\vert}{\vert E_3^{\ast ''}\bm{p}_4^{\ast ''}-E_4^{\ast ''}\bm{p}_3^{\ast ''}\vert}.
\end{eqnarray}
In the normal case that both kinetic and canonical energies (momentums) are conserved 
in a process, the $F^{\prime}$ and $F^{\prime\prime}$ frames are the same, and the detailed balance relation becomes 
\begin{equation}
\sigma_{34} = \frac{\sigma_{12}}{2(1+\delta_{12})}\frac{2\pi p_1^{\ast 2}}{p_4^{\ast}\int dm \mathcal{A}(m)p_4^{\ast}(m)}, 
\end{equation}
where $p_1^*$  and $p_4^*$ are particle $kinetic$ momenta in the center of mass frame $\bm{p}_1^*+\bm{p}_2^*=\bm{p}_3^*+\bm{p}_4^*=0$.  

The cross section for $N+\pi \rightarrow \Delta$ is given by
\begin{eqnarray}
\sigma_{N\pi\to\Delta} &= &\int\frac{dm}{2\pi}\mathcal{A}({m})
\int\frac{d^3\bm{p}^*_{\Delta}}{(2\pi)^3}\frac{1}{2E_{\Delta}^*}
\frac{\vert \mathcal{M}\vert^2_{N\pi\rightarrow\Delta}}{4E_N^*E_{\pi}\vert v_N-v_{\pi}\vert} \notag \\
& & \times(2\pi)^4\delta^4(p_N+p_{\pi}-p_{\Delta}), \notag\\
& =& \frac{\mathcal{A}(m_{\Delta})}{8\sqrt{m_{\Delta}^{*2}+\bm{p}_{\Delta}^{*2}}}\frac{\vert \mathcal{M}\vert^2_{N\pi\rightarrow\Delta}}{E_N^*E_{\pi}\vert v_N-v_{\pi}\vert}
\end{eqnarray}
where $\bm{v}_N=\bm{p}_N^*/E_N^{\ast}$ and  $\bm{v}_{\pi}=\bm{p}_{\pi}^*/E_{\pi}^{\ast}$. In the center of mass frame of $\bm{p}_N^*+\bm{p}_{\pi}=0$, since $\vert v_N-v_{\pi}\vert=\vert \frac{\bm{p}_N^*}{E_N^*}-\frac{\bm{p}_N^*}{E_{\pi}}\vert$ and $\bm{p}_{\Delta}^{*2}=(\bm{\Sigma}_N-\bm{\Sigma}_{\Delta})^2\ll m_{\Delta}^{*2}$, the cross section can be rewritten as 
\begin{equation}
\sigma_{N\pi\to\Delta}  = \frac{\mathcal{A}(m_\Delta)}{8m_{\Delta}^{\ast}}\frac{\vert \mathcal{M}\vert^2_{N\pi\rightarrow\Delta}}{(E_N^*+E_{\pi})\bm{p}_N^*}
\end{equation}

The decay width of $\Delta$ in the frame of $\bm{p}_{\Delta}^*=0$ is given by 
\begin{eqnarray}
\Gamma &=& \frac{1}{2m_{\Delta}^*}\int \frac{d^3\bm{p}^*_N}{(2\pi)^3}\frac{d^3\bm{p}_{\pi}^3}{(2\pi)^3}\frac{\vert \mathcal{M}\vert^2_{\Delta\rightarrow N\pi} }{4E_{N}^*E_{\pi}\vert v_N-v_{\pi}\vert} \notag \\
& & \times(2\pi)^4\delta^4(p_N+p_{\pi}-p_{\Delta}).
\end{eqnarray}
Calculating the integral in the frame of $\bm{p}_N^{\ast}+\bm{p}_{\pi}=0$, we obtain
\begin{equation}
\Gamma = \frac{1}{8\pi m_{\Delta}^{\ast}}\frac{p_N^{*}}{E_N^{*}+E_{\pi}}
\vert\mathcal{M}\vert^2_{\Delta\rightarrow N\pi}. 
\end{equation}
Using the relation $2\vert\mathcal{M}\vert^2_{\Delta\rightarrow N\pi}=
\vert \mathcal{M}\vert^2_{N\pi\rightarrow\Delta}$ leads to
\begin{equation}
\sigma_{N\pi\to\Delta} = \frac{2\pi}{p_N^{*2}}\mathcal{A}(m_{\Delta})\Gamma(m_{\Delta}).
\end{equation}

We note that the above detailed balance relations do not depend on the form of $\Delta$ spectral function $\mathcal{A}(m)$ and $\Delta$ decay width $\Gamma (m)$ used in the derivation. In principle, in-medium potentials also affect $\mathcal{A}(m)$ and $\Gamma (m)$~\cite{Zhang2017}. In the present study, we use those in vacuum for simplicity.

\subsection{Scattering cross sections}

For baryon-baryon elastic scattering, we use the total cross section
\begin{multline}
\sigma_{BB\rightarrow BB}^{\mathrm{elastic}} (\mathrm{mb}) \\
= \left\{ 
\begin{aligned}
& 55,&\sqrt{s}<1.8993 \mathrm{GeV},\\
& 20+\frac{35}{1+100(\sqrt{s}-1.8993)},&\sqrt{s}\geqslant 1.8993 \mathrm{GeV}, 
\end{aligned}
\right.
\end{multline}
and the differential cross section
\begin{equation}
\frac{d\sigma_{BB\rightarrow BB}^{\mathrm{elastic}}}{dt}\sim
\mathrm{exp}\left[ 
\frac{6\{3.65(\sqrt{s}-1.866\}^6}{1+\{3.65(\sqrt{s}-1.866\}^6}t
\right],
\end{equation}
as parametrized in Ref.~\cite{Bertsch1988a}. 

For $\Delta$ production from the process $N+N\rightarrow N+\Delta$, we use the cross section predicted by the one-boson exchanged model~\cite{Huber1994}. Based on Eq.~(\ref{Eq:sig12fb}), the mass of produced $\Delta$  resonances is distributed according to 
\begin{eqnarray}
P(m^{\ast}) = \frac{\mathcal{A}(m)p^{\ast}}{\int_{m_{\mathrm{min}}}^{m_{\mathrm{max}}}dm' \mathcal{A}(m')p^*(m')},
\end{eqnarray}
where $m^{\ast}$ is the effective mass of $\Delta$, the $m_{\mathrm{min}}$ and $m_{\mathrm{max}}$ are the minimum and maximum masses of 
$\Delta$ that are allowed to form. In free space, their values are $m_{\mathrm{min}}= m_N+m_{\pi}$ and $m_{\mathrm{max}}= \sqrt{s}-m_N$.

The cross sections for $N+\Delta\rightarrow N+N$　and  $\pi+ N\rightarrow\Delta$ are determined by the detailed balance relations introduced in Sec.~\ref{Sec:DB}. The $\pi+N$ elastic scattering is also included  with  a constant cross section of $20$ mb.

In a scattering or decay, the momentum and energy of final particles are determined by the canonical energy and momentum conservation, that is 
\begin{eqnarray}
\sum_i \bm{p}_i &= & \sum_j\bm{p}_j, \\
\sum_i\left(\sqrt{m_i^{*2}+\bm{p}_i^{*2}}+\Sigma_i^0\right)&=&\sum_j\left(\sqrt{m_j^{*2}+\bm{p}_j^{*2}}+\Sigma_j^0\right),\nonumber\\
\end{eqnarray} 
where $i$ and $j$ run over the particles before and after a reaction, respectively. In the present study, we consider three different cases of energy conservation condition: 1) including both scalar and vector potentials (S+V); 2) including only scalar potential (S); and 3)  without mean-field potentials (free).

\section{Results and discussions}

\begin{table*}[hbt]
\caption{Temperature $T$, neutron ($\mu_n$) and proton ($\mu_p$) chemical potentials, particle numbers, pion-like number $\pi_{\mathrm{like}}$ and effective charge pion ratio $(\pi^-/\pi^+)_{\mathrm{like}}$ in initial and final states in cases `S' and `free' (see text for details) from the thermal model calculations. } \label{Tab:part}
\begin{tabular}{c|ccccccccccccccc}
\hline \hline
& T (MeV)&$\mu_n(\mathrm{MeV})$ &$\mu_p(\mathrm{MeV})$ & N& Z& $\Delta^{++}$ & $\Delta^{+}$ &$\Delta^{0}$ &$\Delta^{-}$ & $\pi^+$ & $\pi^0$ & $\pi^-$ & $\pi_{\mathrm{like}}$ & $(\pi^-/\pi^+)_{\mathrm{like}}$\\
initial &60.0 &902.6 &871.1& 130.9&97.7&1.29 & 2.02&3.16&4.95 & 0.72 &1.22&2.06 &15.42 &3.0 \\
final (S)  &60.3&639.5 &619.6&133.2 &95.7 &1.57 &2.19&3.04&4.24 & 0.90 &1.25 &1.73 &14.91 &2.2 \\
final (free)&51.6&921.5 &902.6   & 138.9&96.2&0.65 & 0.94&1.36&1.96 & 0.43 &0.62&0.89 &6.85 &2.4 \\
\hline\hline
\end{tabular}
\end{table*}

\begin{figure*}[hbt]
\includegraphics[width=0.9\linewidth]{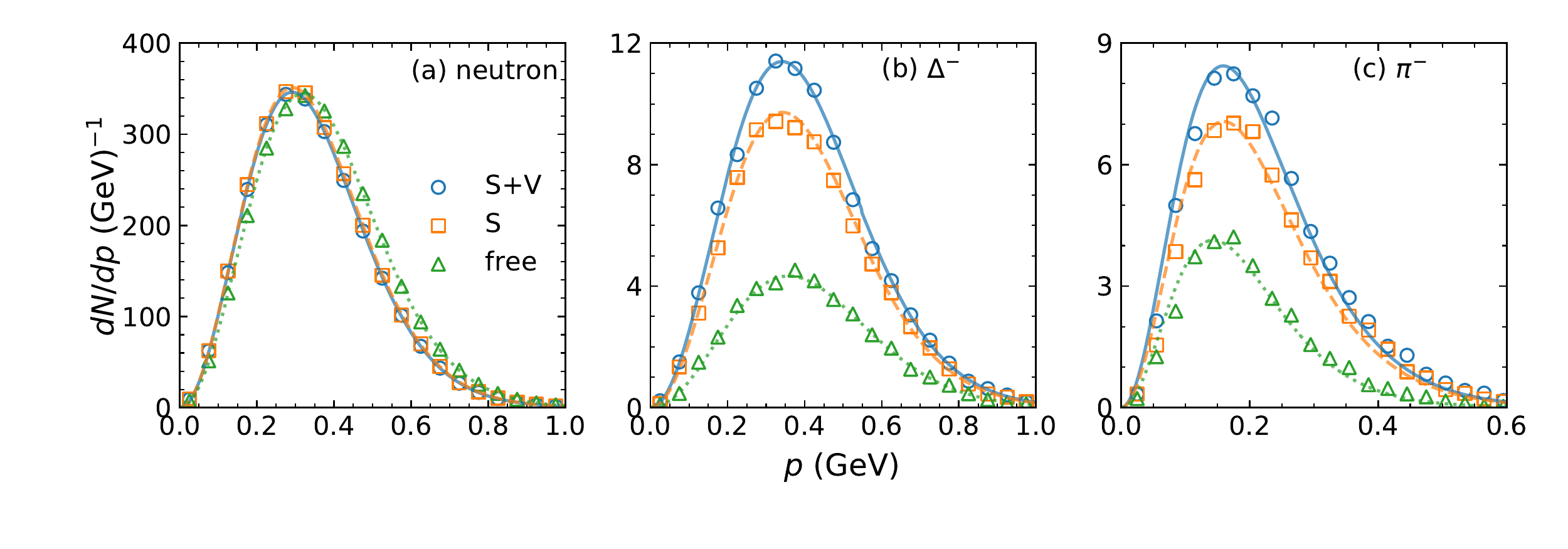}
\caption{(Color online) Momentum distributions of neutron (left panel), $\Delta^-$ (middle panel) and $\pi^-$ (right panel) in a box at $t=40$ fm/$c$. Solid lines are theoretical momentum distributions from the thermal model with the inclusion of both scalar and vector potentials at $T=60$ MeV, dash and dot lines are those in the final sates of `S' and `free' cases (see text for details). Open circles are results from the RVUU model including effects of both scalar and vector potentials in the energy conservation condition of a reaction (S+V), open squares are those including only the scalar potentials (S), and open triangles are those without including any potentials (free).}
\label{Fig:mom}
\end{figure*}

In the present work, we confine all particles in a cubic box of $10\times10\times10$ fm$^{3}$ with periodic boundary conditions. The initial numbers of $N$, $\Delta$ and $\pi$ are determined from the thermal model with fixed temperature $T=60$ MeV, baryon density $\rho=0.24$ fm$^{-3}$, and isovector density $\rho_I=0.096$ fm$^{-3}$, which resemble the conditions of the dense matter, where most pions are produced, in intermediate energy heavy ion collisions~\cite{Xu2013}.  As introduced in Sec.~\ref{Sec:RMF}, the baryon mean-field potentials and densities can be iteratively solved from Eq.~(\ref{chemical}). The resulting mean-field potentials are $g_{\sigma}\sigma = 292.2~\mathrm{MeV}$, $g_{\omega}\omega^0 = 256.6~\mathrm{MeV}$ and $g_{\rho}\rho_3^0= -7.0~\mathrm{MeV}$.  The initial numbers of $N$, $\Delta$ and $\pi$ of various charges are shown in the first row of Table \ref{Tab:part}. For the initial momentum spectra, which are determined by Eqs.~(\ref{Eq:fnpi}) and (\ref{Eq:fdlt}), they are shown by solid lines in Figs.~\ref{Fig:mom} (a), (b) and (c) for neutron, $\Delta^-$ and $\pi^-$, respectively.

Since both the scalar and vector potentials are uniform in the box, all particles move with  constant velocity 
\begin{equation}
{\dot{\bm r}} = \frac{\bm{p}}{\sqrt{m^{*2}+p^2}}
\end{equation}
between scatterings.  Although the scalar potential may change when mean-field potentials are neglected in the energy conservation condition of a scattering or decay process, the changes are small ($\sim 0.1$ MeV) and are thus neglected in the present study by using the same constant value through out the evolution of the system for all scenarios of treating particle scatterings.   

As mentioned in Sec.~\ref{Sec:RMF}, we always use the $\Delta$ decay width in vacuum. Its decay probability in each time step  $dt$ is then given by  
\begin{equation}
P = 1-\mathrm{exp}(-dt\Gamma/\gamma),
\end{equation} 
where $\gamma$ is the Lorentz factor in the frame of $p_{\Delta}^*=0$.  Given that the pion absorption cross section can be as large as about $200~\mathrm{mb}$ in vacuum, we carry out the box calculation by using the partition method of 10 test particles for a physical particle and reducing accordingly all scattering cross sections by 10~\cite{Zhang:1998tj}.  Results shown below are obtained with 400 such events. 

\begin{figure*}[hbt]
\includegraphics[width=0.9\linewidth]{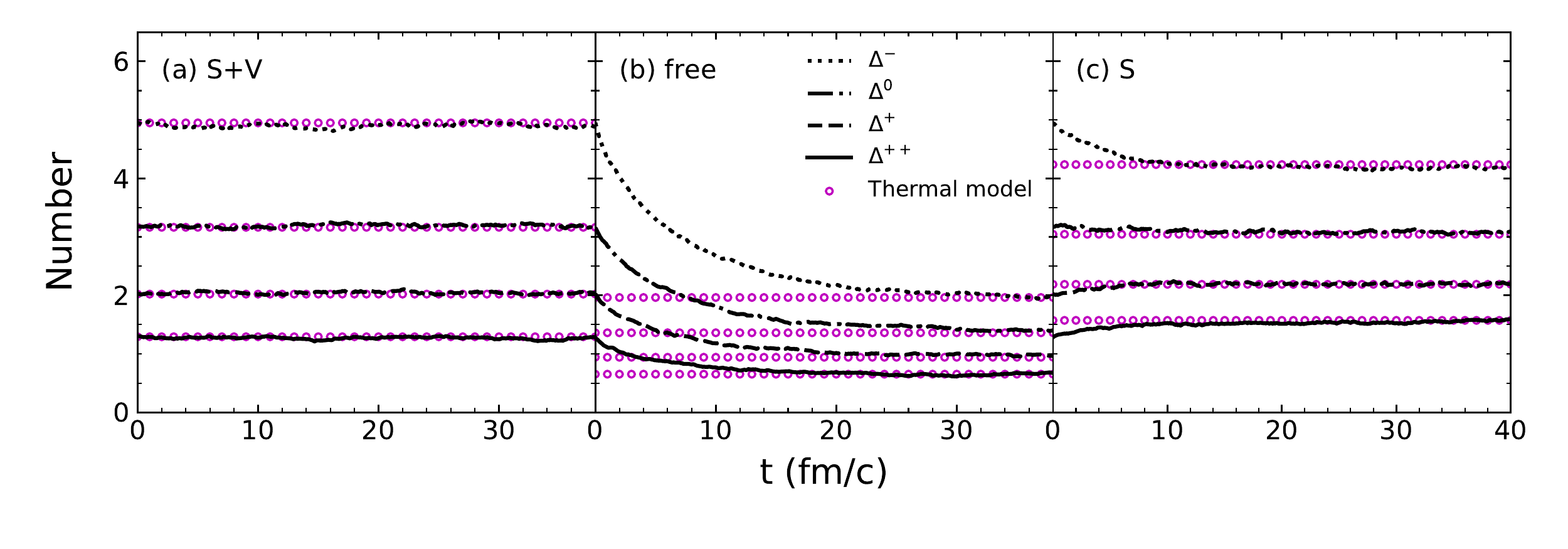}
\caption{(Color online) Time evolutions of the $\Delta$ numbers in box calculations 
in three cases (see text for details). For comparison, the thermal model results 
are shown as open cycles.}\label{Fig:Dlt}
\end{figure*}

\begin{figure*}[hbt]
\includegraphics[width=0.9\linewidth]{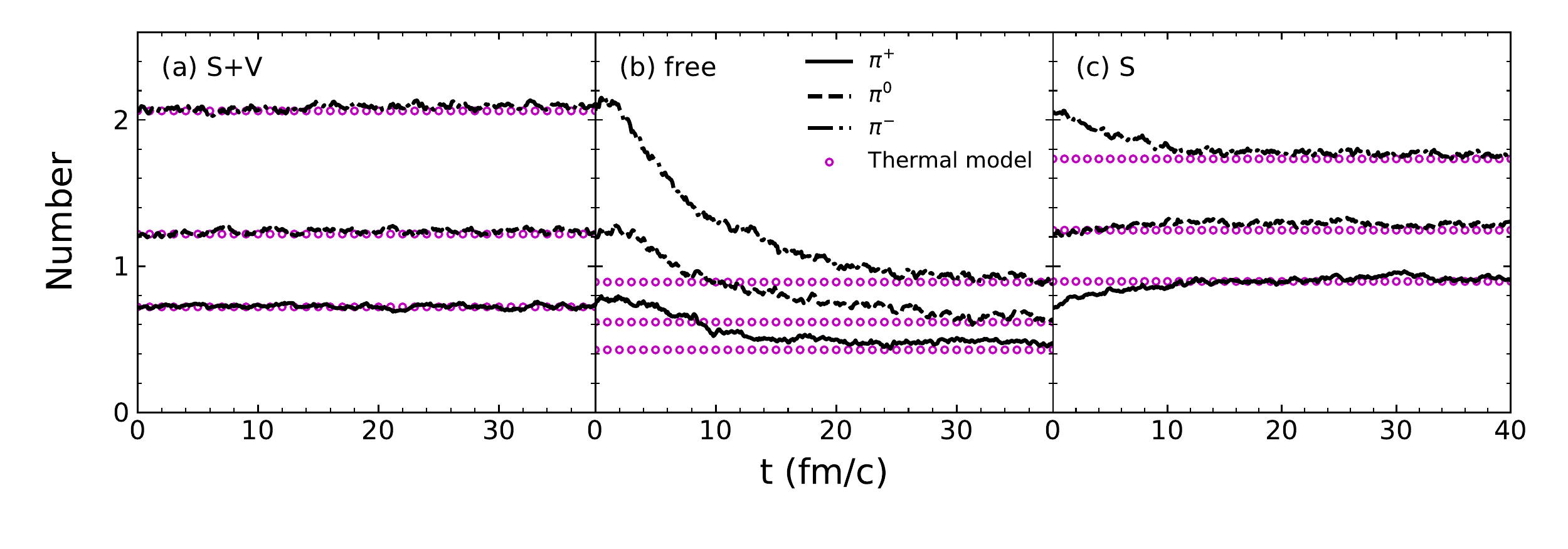}
\caption{(Color online) Simlar to Fig.~\ref{Fig:Dlt} but for the $\pi$ numbers. }\label{Fig:pion}
\end{figure*}

Figs.~\ref{Fig:Dlt} and ~\ref{Fig:pion} show the time evolutions of $\Delta$ and $\pi$ numbers for the three cases of energy conservation condition. As expected, the numbers of $\Delta$ and $\pi$ in the `S+V' case remain almost unchanged except small fluctuations, since their initial numbers are determined from the thermal model with the inclusion of both the scalar and vector potentials. In the `free' case, the numbers of $\Delta$ and $\pi$ decrease with time and reach their equilibrium numbers at $\sim 30~\mathrm{fm}/c$.  For the `S' case, before reaching the equilibrium numbers at $\sim 10 ~\mathrm{fm}/c$, the $\Delta^-$ and $\pi^-$ numbers decrease with time,  the $\Delta^{++}$ and $\pi^+$ numbers increase with time, while the $\Delta^0$,  $\Delta^+$ and $\pi^0$ numbers change only slightly. The final equilibrium numbers in the system for the `S' and `free' cases can also be determined by thermal model calculations using the conditions of energy, baryon density and isospin density conservations. These results are given in the second and third rows of Tab.~\ref{Tab:part}, respectively, and also shown in Figs.~\ref{Fig:Dlt} and~\ref{Fig:pion}, respectively, by open circles.  The RVUU box calculations well reproduce the thermal model results with a deviation less than $\sim 2\%$.  The momentum distributions of neutron, $\Delta^-$ and $\pi^-$ in the box at $t=40~\mathrm{fm}/c$ for the three cases are  shown by open symbols in Figs.~\ref{Fig:mom} (a), (b) and (c), respectively, while corresponding theoretical momentum spectra are exhibited as curves. It is seen that in all  three cases the $N-\Delta-\pi$ system is at the expected thermal equilibrium states.

It is known  in heavy ion collisions that the effective pion number, which includes all pion-like particles, changes very little after the colliding nuclear matter reaches the maximum compression~\cite{Xiong:1993pd,Xu:2017akx}.  In our case, the effective pion number is given by the sum of $\Delta$ resonance and pion numbers, i.e.,  $\pi_{\mathrm{like}}=\pi^-+\pi^0+\pi^++\Delta^{++}+\Delta^++\Delta^0+\Delta^-$, and it is shown in Tab.~\ref{Tab:part}. Also shown in Tab.~\ref{Tab:part} is the effective charged pion ratio $(\pi^-/\pi^+)_{\mathrm{like}}$ defined as 
\begin{eqnarray}
(\pi^-/\pi^+)_{\mathrm{like}}&=& \frac{\pi^-+\Delta^-+\Delta^0/3}{\pi^++\Delta^{++}+\Delta^+/3}.
\end{eqnarray}

Comparing results for the `S+V' and `S' cases, we find that omitting the vector potential in the energy conservation condition slightly decreases the effective pion number $\pi_{\mathrm{like}}$ by $~3.3\%$, but significantly reduces the ratio $(\pi^-/\pi^+)_{\mathrm{like}}$ by $26.7\%$. To understand these results, we note that for particles in thermal and chemical equilibrium, Eq.~(\ref{chemical}) leads to  following relations between the ratio of the numbers of particles in the same isospin multiplet and the chemical potentials of neutron ($\mu_n$) and proton ($\mu_p$),
\begin{equation}
\frac{n}{p}=\frac{\Delta^0}{\Delta^+}=
\left( \frac{\Delta^-}{\Delta^{++}}\right)^{\frac{1}{3}}
=\left( \frac{\pi^-}{\pi^+}\right)^{\frac{1}{2}}=\mathrm{exp}\left(\frac{\mu_n-\mu_p}{T}\right).\\
\end{equation}
Although neglecting the vector potential does not affect the temperature of the system, it changes the chemical potentials of baryons and thus affects the particle ratios.  With both  scalar and vector potentials, the system initially has $\mu_n-\mu_p=31.5~\mathrm{MeV}$. Neglecting the vector potential, this value is reduced to $17.5~\mathrm{MeV}$ but increases to $18.9$ MeV after the system reaches equilibrium.  This explains the significantly smaller value for  the ratio $(\pi^-/\pi^+)_{\mathrm{like}}$ in the `S+V' case than the `S' case. 
 
As to the scalar potential, although it is the same for nucleons and $\Delta$ resonances, its effects on particle numbers are very large.  As shown in the second row of Tab.~\ref{Tab:part}, the $\pi_{\mathrm{like}}$ in the `S' case is about 2.2 times larger than that in the `free' case (third row), while the ratio $(\pi^-/\pi^+)_{\mathrm{like}}$ increases by $9\%$ from the `S' to `free' case. Such large effects are mainly due to the sudden removal of scaler potential, which reduces appreciably the total kinetic energy of the system and thus its temperature.  

The results presented above are based on the assumption that the coupling constants 
of $\Delta$ resonances to mesons are the same as those of nucleons. It has been shown that changing the isovector part of the $\Delta$ potential, such as taking the $\Delta\Delta\rho$ coupling constant  $g_{\rho\Delta}$ to be $3g_{\rho}$, affects the properties of neutron stars~\cite{Cai2015} and pion production in heavy ion collisions~\cite{Li2015, Cozma2017}.  Although using $g_{\rho\Delta}=3g_\rho$ guarantees the conservation of total potential energy  in the reaction $N+N\leftrightarrow N+\Delta$,  the initial and final potentials  in the process $\Delta \leftrightarrow N+\pi$ still differ.  For the hot asymmetric nuclear matter considered in the present study, increasing  $g_{\rho\Delta}$ from $g_{\rho}$ to $3g_{\rho}$ reduces the $\pi_{\mathrm{like}}$ number and the ratio $(\pi^{-}/\pi^{+})_{\mathrm{like}}$ by $3.4\%$ and $17.3\%$,  respectively. The considerable effect on $(\pi^-/\pi^+)_{\mathrm{like}}$  suggests the possibility to study the isovector  potentials of $\Delta$ resonances from pion production in heavy ion 
collisions, as pointed out in Ref.\citep{Li2015}.

Because of the change of the threshold energy of a reaction caused by the baryon potentials,  cross sections for some channels of the reaction $N+N\leftrightarrow N+\Delta$ may increase or decrease. For a system confined in a box as in the present study, changing these cross sections only influences the time for the system to reach a new equilibrium after neglecting the scalar and/or the vector potential, but does not affect the final equilibrium particle numbers. However, in nuclear collisions, where thermal and chemical equilibriums are likely not reached, the change of cross sections can have a large effect on the final pion number and the charged pion ratio~\cite{Song2014}.

\section{summary}

We have employed the relativistic Vlasov-Uheling-Uhlenbeck transport model to study a thermalized $N-\Delta-\pi$ system in a box with periodic boundary conditions. Comparing our results with thermal model calculations, we find that with the inclusion of both baryon scalar and vector potentials in the energy conservation condition for particle production or absorption in scattering and decay processes, our results can well reproduce the equilibrium numbers of particles obtained in thermal model calculations, which verifies the reliability of the RVUU model.  Omitting the vector potentials of baryons in the energy conservation conditions for scattering and decay processes reduces slightly the number of pion-like particles by $~3.3\%$, but  significantly the effective charged pion ratio by $26.7\%$.  Neglecting also the scalar potential further reduces the pion-like particle number by a factor of $\sim 2$, and increases the effective charged pion ratio by about $9\%$.  Our results thus indicate that the correct treatment of the energy conservation condition in scattering and decay processes in transport models is very important for studying pion production in heavy ion collisions at intermediate energies. 

\subsection*{Acknowledgements}

This work was supported by the US Department of Energy under Contract No.\ DE-SC0015266 and the Welch Foundation under Grant No.\ A-1358.

\bibliography{ref.bib}

\begin{thebibliography}{32}
\expandafter\ifx\csname natexlab\endcsname\relax\def\natexlab#1{#1}\fi
\expandafter\ifx\csname bibnamefont\endcsname\relax
  \def\bibnamefont#1{#1}\fi
\expandafter\ifx\csname bibfnamefont\endcsname\relax
  \def\bibfnamefont#1{#1}\fi
\expandafter\ifx\csname citenamefont\endcsname\relax
  \def\citenamefont#1{#1}\fi
\expandafter\ifx\csname url\endcsname\relax
  \def\url#1{\texttt{#1}}\fi
\expandafter\ifx\csname urlprefix\endcsname\relax\def\urlprefix{URL }\fi
\providecommand{\bibinfo}[2]{#2}
\providecommand{\eprint}[2][]{\url{#2}}

\bibitem[{\citenamefont{Li}(2002{\natexlab{a}})}]{Li2002}
\bibinfo{author}{\bibfnamefont{B.-A.} \bibnamefont{Li}},
  \bibinfo{journal}{Phys. Rev. Lett.} \textbf{\bibinfo{volume}{88}},
  \bibinfo{pages}{192701} (\bibinfo{year}{2002}{\natexlab{a}}).

\bibitem[{\citenamefont{Steiner et~al.}(2005)\citenamefont{Steiner, Prakash,
  Lattimer, and Ellis}}]{Steiner2005b}
\bibinfo{author}{\bibfnamefont{A.~W.} \bibnamefont{Steiner}},
  \bibinfo{author}{\bibfnamefont{M.}~\bibnamefont{Prakash}},
  \bibinfo{author}{\bibfnamefont{J.~M.} \bibnamefont{Lattimer}},
  \bibnamefont{and} \bibinfo{author}{\bibfnamefont{P.~J.} \bibnamefont{Ellis}},
  \bibinfo{journal}{Phys. Rep.} \textbf{\bibinfo{volume}{411}},
  \bibinfo{pages}{325} (\bibinfo{year}{2005}).

\bibitem[{\citenamefont{Lattimer and Prakash}(2007)}]{Lattimer2007a}
\bibinfo{author}{\bibfnamefont{J.~M.} \bibnamefont{Lattimer}} \bibnamefont{and}
  \bibinfo{author}{\bibfnamefont{M.}~\bibnamefont{Prakash}},
  \bibinfo{journal}{Phys. Rep.} \textbf{\bibinfo{volume}{442}},
  \bibinfo{pages}{109} (\bibinfo{year}{2007}).

\bibitem[{\citenamefont{Li et~al.}(2008)\citenamefont{Li, Chen, and
  Ko}}]{Li2008}
\bibinfo{author}{\bibfnamefont{B.~A.} \bibnamefont{Li}},
  \bibinfo{author}{\bibfnamefont{L.~W.} \bibnamefont{Chen}}, \bibnamefont{and}
  \bibinfo{author}{\bibfnamefont{C.~M.} \bibnamefont{Ko}},
  \bibinfo{journal}{Phys. Rep.} \textbf{\bibinfo{volume}{464}},
  \bibinfo{pages}{113} (\bibinfo{year}{2008}).

\bibitem[{\citenamefont{Fattoyev et~al.}(2014)\citenamefont{Fattoyev, Newton,
  and Li}}]{Fattoyev2014}
\bibinfo{author}{\bibfnamefont{F.~J.} \bibnamefont{Fattoyev}},
  \bibinfo{author}{\bibfnamefont{W.~G.} \bibnamefont{Newton}},
  \bibnamefont{and} \bibinfo{author}{\bibfnamefont{B.~A.} \bibnamefont{Li}},
  \bibinfo{journal}{Eur. Phys. J. A} \textbf{\bibinfo{volume}{50}},
  \bibinfo{pages}{1} (\bibinfo{year}{2014}).

\bibitem[{\citenamefont{Xiao et~al.}(2009)\citenamefont{Xiao, Li, Chen, Yong,
  and Zhang}}]{Xiao2009}
\bibinfo{author}{\bibfnamefont{Z.}~\bibnamefont{Xiao}},
  \bibinfo{author}{\bibfnamefont{B.~A.} \bibnamefont{Li}},
  \bibinfo{author}{\bibfnamefont{L.~W.} \bibnamefont{Chen}},
  \bibinfo{author}{\bibfnamefont{G.~C.} \bibnamefont{Yong}}, \bibnamefont{and}
  \bibinfo{author}{\bibfnamefont{M.}~\bibnamefont{Zhang}},
  \bibinfo{journal}{Phys. Rev. Lett.} \textbf{\bibinfo{volume}{102}},
  \bibinfo{pages}{062502} (\bibinfo{year}{2009}).

\bibitem[{\citenamefont{Feng and Jin}(2010)}]{Feng2010}
\bibinfo{author}{\bibfnamefont{Z.~Q.} \bibnamefont{Feng}} \bibnamefont{and}
  \bibinfo{author}{\bibfnamefont{G.~M.} \bibnamefont{Jin}},
  \bibinfo{journal}{Phys. Lett. B} \textbf{\bibinfo{volume}{683}},
  \bibinfo{pages}{140} (\bibinfo{year}{2010}).

\bibitem[{\citenamefont{Xie et~al.}(2013)\citenamefont{Xie, Su, Zhu, and
  Zhang}}]{Xie2013}
\bibinfo{author}{\bibfnamefont{W.~J.} \bibnamefont{Xie}},
  \bibinfo{author}{\bibfnamefont{J.}~\bibnamefont{Su}},
  \bibinfo{author}{\bibfnamefont{L.}~\bibnamefont{Zhu}}, \bibnamefont{and}
  \bibinfo{author}{\bibfnamefont{F.~S.} \bibnamefont{Zhang}},
  \bibinfo{journal}{Phys. Lett. B} \textbf{\bibinfo{volume}{718}},
  \bibinfo{pages}{1510} (\bibinfo{year}{2013}).

\bibitem[{SEP()}]{SEP}
\emph{\bibinfo{title}{Symmetry energy project}},
  \bibinfo{howpublished}{\url{https://groups.nscl.msu.edu/hira/sepweb/pages/home.html}}.

\bibitem[{\citenamefont{Xu et~al.}(2010)\citenamefont{Xu, Ko, and Oh}}]{Xu2010}
\bibinfo{author}{\bibfnamefont{J.}~\bibnamefont{Xu}},
  \bibinfo{author}{\bibfnamefont{C.~M.} \bibnamefont{Ko}}, \bibnamefont{and}
  \bibinfo{author}{\bibfnamefont{Y.}~\bibnamefont{Oh}}, \bibinfo{journal}{Phys.
  Rev. C} \textbf{\bibinfo{volume}{81}}, \bibinfo{pages}{2}
  (\bibinfo{year}{2010}).

\bibitem[{\citenamefont{Xu et~al.}(2013)\citenamefont{Xu, Chen, Ko, Li, and
  Ma}}]{Xu2013}
\bibinfo{author}{\bibfnamefont{J.}~\bibnamefont{Xu}},
  \bibinfo{author}{\bibfnamefont{L.~W.} \bibnamefont{Chen}},
  \bibinfo{author}{\bibfnamefont{C.~M.} \bibnamefont{Ko}},
  \bibinfo{author}{\bibfnamefont{B.~A.} \bibnamefont{Li}}, \bibnamefont{and}
  \bibinfo{author}{\bibfnamefont{Y.~G.} \bibnamefont{Ma}},
  \bibinfo{journal}{Phys. Rev. C} \textbf{\bibinfo{volume}{87}},
  \bibinfo{pages}{1} (\bibinfo{year}{2013}).

\bibitem[{\citenamefont{Cozma}(2017)}]{Cozma2017}
\bibinfo{author}{\bibfnamefont{M.~D.} \bibnamefont{Cozma}},
  \bibinfo{journal}{Phys. Rev. C} \textbf{\bibinfo{volume}{95}},
  \bibinfo{pages}{014601} (\bibinfo{year}{2017}).

\bibitem[{\citenamefont{Zhang and Ko}(2017)}]{Zhang2017}
\bibinfo{author}{\bibfnamefont{Z.}~\bibnamefont{Zhang}} \bibnamefont{and}
  \bibinfo{author}{\bibfnamefont{C.~M.} \bibnamefont{Ko}},
  \bibinfo{journal}{Phys. Rev. C} \textbf{\bibinfo{volume}{95}},
  \bibinfo{pages}{064604} (\bibinfo{year}{2017}).

\bibitem[{\citenamefont{Li}(2015)}]{Li2015}
\bibinfo{author}{\bibfnamefont{B.~A.} \bibnamefont{Li}},
  \bibinfo{journal}{Phys. Rev. C} \textbf{\bibinfo{volume}{92}},
  \bibinfo{pages}{034603} (\bibinfo{year}{2015}).

\bibitem[{\citenamefont{Ferini et~al.}(2005)\citenamefont{Ferini, Colonna,
  Gaitanos, and {Di Toro}}}]{Ferini2005}
\bibinfo{author}{\bibfnamefont{G.}~\bibnamefont{Ferini}},
  \bibinfo{author}{\bibfnamefont{M.}~\bibnamefont{Colonna}},
  \bibinfo{author}{\bibfnamefont{T.}~\bibnamefont{Gaitanos}}, \bibnamefont{and}
  \bibinfo{author}{\bibfnamefont{M.}~\bibnamefont{{Di Toro}}},
  \bibinfo{journal}{Nucl. Phys. A} \textbf{\bibinfo{volume}{762}},
  \bibinfo{pages}{147} (\bibinfo{year}{2005}).

\bibitem[{\citenamefont{Song and Ko}(2015)}]{Song2014}
\bibinfo{author}{\bibfnamefont{T.}~\bibnamefont{Song}} \bibnamefont{and}
  \bibinfo{author}{\bibfnamefont{C.~M.} \bibnamefont{Ko}},
  \bibinfo{journal}{Phys. Rev. C} \textbf{\bibinfo{volume}{91}},
  \bibinfo{pages}{014901} (\bibinfo{year}{2015}).

\bibitem[{\citenamefont{Wei et~al.}(2014)\citenamefont{Wei, Li, Xu, and
  Chen}}]{Wei2014}
\bibinfo{author}{\bibfnamefont{G.-F.} \bibnamefont{Wei}},
  \bibinfo{author}{\bibfnamefont{B.-A.} \bibnamefont{Li}},
  \bibinfo{author}{\bibfnamefont{J.}~\bibnamefont{Xu}}, \bibnamefont{and}
  \bibinfo{author}{\bibfnamefont{L.-W.} \bibnamefont{Chen}},
  \bibinfo{journal}{Phys. Rev. C} \textbf{\bibinfo{volume}{90}},
  \bibinfo{pages}{014610} (\bibinfo{year}{2014}).

\bibitem[{\citenamefont{Li et~al.}(2015)\citenamefont{Li, Guo, and
  Shi}}]{Li2015a}
\bibinfo{author}{\bibfnamefont{B.~A.} \bibnamefont{Li}},
  \bibinfo{author}{\bibfnamefont{W.~J.} \bibnamefont{Guo}}, \bibnamefont{and}
  \bibinfo{author}{\bibfnamefont{Z.}~\bibnamefont{Shi}},
  \bibinfo{journal}{Phys. Rev. C} \textbf{\bibinfo{volume}{91}},
  \bibinfo{pages}{1} (\bibinfo{year}{2015}).

\bibitem[{\citenamefont{Cai et~al.}(2015)\citenamefont{Cai, Fattoyev, Li, and
  Newton}}]{Cai2015}
\bibinfo{author}{\bibfnamefont{B.~J.} \bibnamefont{Cai}},
  \bibinfo{author}{\bibfnamefont{F.~J.} \bibnamefont{Fattoyev}},
  \bibinfo{author}{\bibfnamefont{B.~A.} \bibnamefont{Li}}, \bibnamefont{and}
  \bibinfo{author}{\bibfnamefont{W.~G.} \bibnamefont{Newton}},
  \bibinfo{journal}{Phys. Rev. C} \textbf{\bibinfo{volume}{92}},
  \bibinfo{pages}{015802} (\bibinfo{year}{2015}).

\bibitem[{\citenamefont{Li}(2002{\natexlab{b}})}]{Li2002b}
\bibinfo{author}{\bibfnamefont{B.-A.} \bibnamefont{Li}},
  \bibinfo{journal}{Nucl. Phys. A} \textbf{\bibinfo{volume}{708}},
  \bibinfo{pages}{365} (\bibinfo{year}{2002}{\natexlab{b}}).

\bibitem[{\citenamefont{Cozma}(2016)}]{Cozma2016}
\bibinfo{author}{\bibfnamefont{M.~D.} \bibnamefont{Cozma}},
  \bibinfo{journal}{Phys. Lett. B} \textbf{\bibinfo{volume}{753}},
  \bibinfo{pages}{166} (\bibinfo{year}{2016}).

\bibitem[{\citenamefont{Ko et~al.}(1987)\citenamefont{Ko, Li, and
  Wang}}]{Ko1987}
\bibinfo{author}{\bibfnamefont{C.~M.} \bibnamefont{Ko}},
  \bibinfo{author}{\bibfnamefont{Q.}~\bibnamefont{Li}}, \bibnamefont{and}
  \bibinfo{author}{\bibfnamefont{R.}~\bibnamefont{Wang}},
  \bibinfo{journal}{Phys. Rev. Lett.} \textbf{\bibinfo{volume}{59}},
  \bibinfo{pages}{1084} (\bibinfo{year}{1987}).

\bibitem[{\citenamefont{Ko and Li}(1988)}]{Ko1988}
\bibinfo{author}{\bibfnamefont{C.~M.} \bibnamefont{Ko}} \bibnamefont{and}
  \bibinfo{author}{\bibfnamefont{Q.}~\bibnamefont{Li}}, \bibinfo{journal}{Phys.
  Rev. C} \textbf{\bibinfo{volume}{37}}, \bibinfo{pages}{2270}
  (\bibinfo{year}{1988}).

\bibitem[{\citenamefont{Ko and Li}(1996)}]{Ko1996}
\bibinfo{author}{\bibfnamefont{C.~M.} \bibnamefont{Ko}} \bibnamefont{and}
  \bibinfo{author}{\bibfnamefont{G.~Q.} \bibnamefont{Li}},
  \bibinfo{journal}{Journal of Physics G: Nuclear and Particle Physics}
  \textbf{\bibinfo{volume}{22}}, \bibinfo{pages}{1673} (\bibinfo{year}{1996}).

\bibitem[{\citenamefont{Xu et~al.}(2016)}]{Xu:2016lue}
\bibinfo{author}{\bibfnamefont{J.}~\bibnamefont{Xu}} \bibnamefont{et~al.},
  \bibinfo{journal}{Phys. Rev. C} \textbf{\bibinfo{volume}{93}},
  \bibinfo{pages}{044609} (\bibinfo{year}{2016}).

\bibitem[{\citenamefont{Liu et~al.}(2002)\citenamefont{Liu, Greco, Baran,
  Colonna, and {Di Toro}}}]{Liu2002}
\bibinfo{author}{\bibfnamefont{B.}~\bibnamefont{Liu}},
  \bibinfo{author}{\bibfnamefont{V.}~\bibnamefont{Greco}},
  \bibinfo{author}{\bibfnamefont{V.}~\bibnamefont{Baran}},
  \bibinfo{author}{\bibfnamefont{M.}~\bibnamefont{Colonna}}, \bibnamefont{and}
  \bibinfo{author}{\bibfnamefont{M.}~\bibnamefont{{Di Toro}}},
  \bibinfo{journal}{Phys. Rev. C} \textbf{\bibinfo{volume}{65}},
  \bibinfo{pages}{045201} (\bibinfo{year}{2002}).

\bibitem[{\citenamefont{Kitazoe et~al.}(1986)\citenamefont{Kitazoe, Sano, Toki,
  and Nagamiya}}]{Kitazoe1986}
\bibinfo{author}{\bibfnamefont{Y.}~\bibnamefont{Kitazoe}},
  \bibinfo{author}{\bibfnamefont{M.}~\bibnamefont{Sano}},
  \bibinfo{author}{\bibfnamefont{H.}~\bibnamefont{Toki}}, \bibnamefont{and}
  \bibinfo{author}{\bibfnamefont{S.}~\bibnamefont{Nagamiya}},
  \bibinfo{journal}{Phys. Lett. B} \textbf{\bibinfo{volume}{166}},
  \bibinfo{pages}{35} (\bibinfo{year}{1986}).

\bibitem[{\citenamefont{Bertsch and {Das Gupta}}(1988)}]{Bertsch1988a}
\bibinfo{author}{\bibfnamefont{G.~F.} \bibnamefont{Bertsch}} \bibnamefont{and}
  \bibinfo{author}{\bibfnamefont{S.}~\bibnamefont{{Das Gupta}}},
  \bibinfo{journal}{Phys. Rep.} \textbf{\bibinfo{volume}{160}},
  \bibinfo{pages}{189} (\bibinfo{year}{1988}).

\bibitem[{\citenamefont{Huber and Aichelin}(1994)}]{Huber1994}
\bibinfo{author}{\bibfnamefont{S.}~\bibnamefont{Huber}} \bibnamefont{and}
  \bibinfo{author}{\bibfnamefont{J.}~\bibnamefont{Aichelin}},
  \bibinfo{journal}{Nucl. Phys. A} \textbf{\bibinfo{volume}{573}},
  \bibinfo{pages}{587} (\bibinfo{year}{1994}).

\bibitem[{\citenamefont{Zhang et~al.}(1998)\citenamefont{Zhang, Gyulassy, and
  Pang}}]{Zhang:1998tj}
\bibinfo{author}{\bibfnamefont{B.}~\bibnamefont{Zhang}},
  \bibinfo{author}{\bibfnamefont{M.}~\bibnamefont{Gyulassy}}, \bibnamefont{and}
  \bibinfo{author}{\bibfnamefont{Y.}~\bibnamefont{Pang}},
  \bibinfo{journal}{Phys. Rev. C} \textbf{\bibinfo{volume}{58}},
  \bibinfo{pages}{1175} (\bibinfo{year}{1998}).

\bibitem[{\citenamefont{Xiong et~al.}(1993)\citenamefont{Xiong, Ko, and
  Koch}}]{Xiong:1993pd}
\bibinfo{author}{\bibfnamefont{L.}~\bibnamefont{Xiong}},
  \bibinfo{author}{\bibfnamefont{C.~M.} \bibnamefont{Ko}}, \bibnamefont{and}
  \bibinfo{author}{\bibfnamefont{V.}~\bibnamefont{Koch}},
  \bibinfo{journal}{Phys. Rev. C} \textbf{\bibinfo{volume}{47}},
  \bibinfo{pages}{788} (\bibinfo{year}{1993}).

\bibitem[{\citenamefont{Xu and Ko}(2017)}]{Xu:2017akx}
\bibinfo{author}{\bibfnamefont{J.}~\bibnamefont{Xu}} \bibnamefont{and}
  \bibinfo{author}{\bibfnamefont{C.~M.} \bibnamefont{Ko}},
  \bibinfo{journal}{Phys. Lett. B} \textbf{\bibinfo{volume}{772}},
  \bibinfo{pages}{290} (\bibinfo{year}{2017}).

\end{thebibliography}
\end{document}